\documentclass[
reprint,
amsmath,amssymb,
aps,
prl
]{revtex4-2}
\pdfoutput=1

\usepackage{MnSymbol} 
\usepackage[english]{babel}
\usepackage{hyperref}
\hypersetup{colorlinks=true,
            linkcolor=black,
            citecolor=black,
            urlcolor=blue}
\usepackage{amssymb}
\usepackage{courier}
\usepackage{dsfont}

\usepackage{physics}
\usepackage{wasysym}

\usepackage{graphicx}
\usepackage{dcolumn}
\usepackage{bm}

\begin{document}

\title{Reduced Basis Method for Driven-Dissipative Quantum Systems}

\author{Hans Christiansen$^{1}$}
\author{Virgil V. Baran$^{2,3\,}$}
\author{Jens Paaske$^{1}$}
\affiliation{$^{1}$Center for Quantum Devices, Niels Bohr Institute, University of Copenhagen, 2100 Copenhagen, Denmark}
\affiliation{$^{2}$Faculty of Physics, University of Bucharest, 405 Atomi\c stilor, RO-077125, Bucharest-M\u agurele, Romania}
\affiliation{$^{3}$"Horia Hulubei" National Institute of Physics and Nuclear Engineering, 30 Reactorului, RO-077125, Bucharest-M\u agurele, Romania}

\date{May 8, 2025}

\begin{abstract}

Reduced basis methods provide an efficient way of mapping out phase diagrams of strongly correlated many-body quantum systems. The method relies on using the exact solutions at select parameter values to construct a low-dimensional basis, from which observables can be efficiently and reliably computed throughout the parameter space. Here we show that this method can be generalized to driven-dissipative Markovian systems allowing efficient calculations of observables in the transient and steady states. A subsequent distillation of the reduced basis vectors according to their explained variances allows for an unbiased exploration of the most pronounced parameter dependencies indicative of phase boundaries in the thermodynamic limit.  

\end{abstract}

\maketitle

Driven-dissipative quantum many-body systems are currently being explored in many different experimental platforms, revealing a plethora of different nonequilibrium steady states and transitions between them~\cite{Fink2018Apr,  
Ma2017Apr, Sidler2017Mar, Delteil2019Mar, Harder2018Sep, Yu2024Apr, lu_jan2015,Beaulieu2025Mar, fitzpatrickfeb2017, Finkjan2017, Wu2024Sep, Ding2024Mar}. The rich phenomenology of these steady states is a result of the competition between the unitary dynamics of the system and the drive and dissipation arising from a coupling to an environment~\cite{diehl_quantum_2008, prosen2011sep, lee_unconventional_2013, Lee2013Dec, Sieberer2013May, Dutta2019Dec, Sa2020Apr}. For Markovian systems, the time evolution of an open quantum system can be described by the Lindblad master equation~\cite{breuer_theory_2006}. The direct numerical solution of this equation suffers from an exponential scaling of the Hilbert space dimension with system size. While several numerical methods have been invented to mitigate this unfavorable scaling~\cite{weimer_simulation_2021}, the efficient and reliable simulation of driven-dissipative quantum many-body systems remains a daunting theoretical challenge. This is especially true if one is interested in tracking the steady state when varying coupling constants, energy levels, and other system parameters. 
This problem has recently been investigated in Ref.~\onlinecite{melo_variational_2025}, using variational perturbation theory. In this work, we take a different approach employing the reduced basis method (RBM).

For closed quantum systems, the RBM has been employed to understand the low energy states of a Hamiltonian depending on a set of parameters~\cite{Herbst2022Apr, baran2023, Brehmer2023Apr, Duguet2024Aug}. This is done by reconstructing the state space as a low-dimensional interpolation of a selection of exact states calculated at suitably chosen points in parameter space.
In this work, we extend this idea to encompass driven-dissipative quantum systems, showing how to construct a reduced basis surrogate model which efficiently captures the same system dynamics as the full model, while being numerically much more tractable. With this method, one may conduct fast and elaborate parameter scans, which in turn allows for a subsequent principal component analysis (PCA) and a final distillation of the reduced basis according to explained variances. 
As we demonstrate in two select examples, this procedure provides an effective low-dimensional description of the system, in terms of basis vectors corresponding to the observables that capture the most dramatic changes in a given parameter range. 


\textit{Lindblad Master Equation.} We shall restrict our attention to Markovian systems comprising a quantum system described by a Hilbert space $\mathcal H$ of finite dimension $d_{\mathcal H}$ coupled to an environment such that the time evolution of the reduced density matrix of the system, $\rho(t)$, is described by the time-local Lindblad Master Equation (LME)~\cite{breuer_theory_2006}
\begin{equation}\label{eq:lme}
    \partial_t \rho= -i [H,\rho] + \sum_n \left[L_n\rho L_n^\dagger -\frac{1}{2}\{L_n^\dagger L_n,\rho\}\right].
\end{equation}
Here $H$ is the Lamb-shifted system Hamiltonian and $L_n$ are the jump operators describing the incoherent dynamics induced by the coupling to the environment. The right hand side of Eq.~\eqref{eq:lme} defines the Liouvillian superoperator ${\mathcal L}$, acting on the doubled Hilbert space $\mathcal H \otimes \mathcal H$ and the LME is conveniently written as
$\partial_t \vert \rho \rrangle=  \mathcal{L}\vert \rho \rrangle$, 
where $\vert \rho \rrangle$ now refers to the vectorized version of the reduced density matrix of dimension $d_{\mathcal H}^2$. $\mathcal L$ is always guaranteed to have a zero eigenvalue with the steady-state density matrix corresponding to the right eigenvector $\vert \rho_{\rm ss}\rrangle$. In the absence of special symmetries, this steady state is uniquely determined, meaning that any initial state will flow towards the same steady state~\cite{albert_symmetries_2014}. 

\textit{Reduced basis Methodology.} We consider the Liouvillian $\mathcal L = \mathcal L(\xi)$ depending on a set of parameters $\xi = (\xi_1,\xi_2..)\in \Xi$ where $\Xi$ defines the parameter range of interest. We will now describe the RBM to efficiently compute the steady state over the entire parameter range. As described in the Supplemental Material~\cite{si}, this can easily be generalized to also compute the decaying modes. Henceforth, we suppress the ${\rm ss}$ subscript on the steady-state density matrix. We consider a typical problem for which a numerically expensive solver can be used to compute the exact steady state density matrix, $\rho^{\mathrm{ex}}(\xi)$, at a given parameter point $\xi$. The central RBM idea is to construct a low-dimensional ($d_{\mathrm{rb}}$) {\it reduced basis} $\{\rho_{\mu}\}$ from a number of well-chosen exact solutions, such that the steady state density matrix at {\it any} $\xi$ may be accurately expressed as~
\begin{align}\label{eq:rhorbexpansion}
\rho^{\rm rb}(\xi)&=\sum_{\mu=1}^{d_{\mathrm{rb}}}\alpha_\mu(\xi) \rho_\mu.
\end{align}
While Eq.~\eqref{eq:rhorbexpansion} is formally similar to the expansion for closed quantum systems~\cite{Duguet2024Aug}, the interpretation is very different. For closed quantum systems, this would correspond to a coherent superposition of state vectors, whereas Eq.~\eqref{eq:rhorbexpansion} represents a statistical ensemble of density matrices.
Since the exact density matrix is a zero-vector of $\hat{\mathcal{L}}(\xi)$, the accuracy with which $\rho^{\rm rb}(\xi)\approx \rho^{\rm ex}(\xi)$ may be conveniently assessed by a {\it residual} defined as 
\begin{equation}\label{eq:res0}
    \text{Res}(\xi)\equiv\min_{\alpha}\frac{\llangle \rho^{\mathrm{rb}}(\xi) \vert\hat {\mathcal L}^\dag(\xi) \hat{\mathcal L}(\xi)\vert\rho^{\mathrm{rb}}(\xi)\rrangle}{\llangle \rho^{\mathrm{rb}}(\xi)\vert \rho^{\mathrm{rb}}(\xi)\rrangle}.
\end{equation}
Here we use the Hilbert-Schmidt inner product $\llangle \rho_1 \vert \rho_2 \rrangle  = \Tr{\rho_1^\dag \rho_2}$ with its corresponding norm, $\norm{\rho}\equiv \llangle \rho |\rho\rrangle^{1/2}$. 

\begin{figure}
\includegraphics[width=\linewidth]{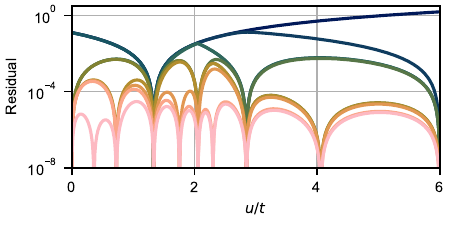}
\caption{Residual for the Fermi-Hubbard Chain with $L=8$ sites and the parameter $\xi=V\in[0,6t]$. The residual vanishes exactly at the sample points while the residual between snapshots gradually becomes smaller.}
\label{fig:1dres}
\end{figure}

In practice, one generates the basis iteratively. Starting from an orthonormal basis $\{\rho_\mu\}$, the residual is simply computed as the smallest eigenvalue of the positive semi-definite matrix $(\hat {\mathcal L}^\dag \hat{\mathcal L})_{\mu\nu}(\xi)=\llangle \rho_\mu \vert \hat {\mathcal L}^\dag(\xi) \hat{\mathcal L}(\xi)\vert \rho_\nu\rrangle$, with the expansion coefficients $\alpha_\mu(\xi)$ being the corresponding eigenvector~\cite{horn2012matrix}, normalized such that the trace of $\rho^{\rm rb}(\xi)$ is one. Following a greedy optimization method~\cite{sarkar_self-learning_2022}, we obtain the next parameter point, at which to use the exact solver, by maximizing the residual over the entire parameter space. Adding this solution to the reduced basis and performing a subsequent orthonormalization completes the iterative step. 
This construction is terminated at some dimension, $d_{\mathrm{rb}}$, for which the maximum residual over the entire parameter range drops below a desired threshold (cf. Fig.~\ref{fig:1dres}). 
The evaluation of the residual across the parameter domain is particularly efficient for Liouvillians that admit an affine decomposition, 
\begin{equation}
    \hat{\mathcal{L}}(\xi)=\sum_\ell f_\ell(\xi)\, \hat{\mathcal{L}}_\ell,
\end{equation}
with the $f_\ell(\xi)$ functions dictating the full parameter
dependence of the system’s Liouvillian, $\hat{\mathcal{L}}(\xi)$, while $\hat{\mathcal{L}}_\ell$ is independent of $\xi$. We can therefore obtain $\hat{\mathcal L}^\dag(\xi) \hat{\mathcal L}(\xi)$ as $(\hat{\mathcal L}^\dag \hat{\mathcal L})_{\mu\nu}(\xi)=\sum_{\ell m} f_\ell^*(\xi)f_m(\xi)\,\llangle\rho_\mu|\hat{\mathcal{L}}_\ell^\dagger\hat{\mathcal{L}}_m|\rho_\nu\rrangle$, the latter matrix elements being evaluated only once. It should be noted that it is not a priori clear how the accuracy of expectation values, computed in the reduced basis, scales with the residual in Eq.~\eqref{eq:res0}, but as shown in the Supplementary Material~\cite{si} the error for an observable $O$ is bounded at least as
$\norm{\delta \expval{O}}/\norm{O}\leq(1+\sqrt{d_{\mathcal H}}) \norm{\rho^{\mathrm{rb}}}\sqrt{\mathrm{Res}(\xi)/\Gamma}$, where $\Gamma$ denotes the spectral gap of $\hat{\mathcal L}^\dag(\xi) \hat{\mathcal L}(\xi)$. In the examples we have considered, this bound is satisfied to many orders of magnitude.

Due to the low dimensionality of the reduced basis, it is now numerically tractable to calculate averages over the entire parameter space, denoted by $\langle \cdot \rangle_\Xi$, and thereby the covariance matrix
\begin{equation}
    C^\Xi_{\mu\nu} = \left\langle \left(\alpha_\mu(\xi) - \langle \alpha_\mu\rangle_\Xi\right) \left(\alpha_\nu(\xi) - \langle \alpha_\nu\rangle_\Xi\right)\right\rangle_{\Xi}.
\end{equation}
One may now transform to a basis in which $C^\Xi$ is diagonal. In this basis, Eq.~\eqref{eq:rhorbexpansion} is written as 
\begin{equation}
    \rho^{\rm rb}(\xi) = \rho_0 + \sum_{\mu=1}^{d_{\mathrm{rb}}} \alpha_\mu^{\rm pc} (\xi) \rho_\mu^{\rm pc},
\end{equation}
with $\rho_0 =\langle \rho^{\rm rb}(\xi)\rangle_\Xi$ and where $\alpha_\mu^{\rm pc}(\xi)$ is the $\mu$'th principal component (PC) fulfilling $\langle \alpha_\mu^{\rm pc}\alpha_\nu^{\rm pc}\rangle_\Xi=\delta_{\mu\nu}c_{\mu}$ with $c_{\mu}$ denoting the explained variances, i.e. the eigenvalues of $C^{\Xi}$~\cite{jolliffe_pca_2016}. 
The corresponding $|\rho^{\mathrm{pc}}_\mu\rrangle$ will be referred to as the PC basis vectors. The expectation value of an observable, $O$, is now computed as  
\begin{equation}\label{eq:pctoobs}
    \expval{O}_{\rho^{\rm rb}(\xi)} = \llangle O \vert \rho^{\rm rb}(\xi)\rrangle =\expval{O}_{\rho_0} + \sum_{\mu=1}^{d_{\rm rb}} \alpha^{\rm pc}_\mu(\xi)\llangle O \vert \rho^{\rm pc}_\mu \rrangle,
\end{equation}
where the overlaps, $\llangle O \vert \rho^{\rm pc}_\mu \rrangle$, relate the parameter dependence of an observable to that of the PCs, $\alpha_{\mu}^{\mathrm{pc}}(\xi)$. Using $O=\rho_\mu^{\rm pc}$, the explained variances can be written as 
\begin{equation}\label{eq:cmudef}
    c_\mu = \expval{\expval{\rho_\mu^{\rm pc}}_{\rho^{\rm rb}(\xi)}^2}_\Xi - \expval{\rho_\mu^{\rm pc}}^2_{\rho_0} = \expval{\llangle \rho_\mu^{\rm pc}\vert \rho^{\rm rb}(\xi)-\rho_0\rrangle^2}_{\Xi}.
\end{equation}
The first equality provides an interpretation of $c_{\mu}$ as the spread in the observable $\rho_\mu^{\rm pc}$ across the parameter range, and the second expresses this as a projected state variance. In the following, we will refer to the normalized eigenvalues, $\lambda_{\mu}=c_{\mu}/(\sum_{\nu}c_{\nu})$, as the explained variance ratios. 

Working in the PC basis serves a dual purpose. Firstly, the ordering according to the explained variance ratios allows the assessment of which parameter directions are truly decisive for the system behavior. Secondly, it allows for a classification of the basis vectors into irreducible representations of a group of symmetry transformations, $g$, acting on both the Liouville and parameter space as  
\begin{align}\label{eq:symtransform}
\hat{\mathcal L}(\xi)\stackrel{g}{\longmapsto} \hat{\mathcal U}^\dag(g) \hat{\mathcal L}(u^{-1}(g)\xi)\hat{\mathcal U}(g).
\end{align}
where $\hat{\mathcal U}(g)$ is the symmetry representation in Liouville space and $u(g)$, the representation in parameter space. The set of symmetries $\{g\}$ which leave the Liouvillian invariant, according to Eq.~\eqref{eq:symtransform}, also forms the symmetry group of the corresponding covariance matrix, provided that the parameter range $\Xi$ is invariant under the symmetry group (cf. Supplementary Material~\cite{si}). The PCs, $\alpha_\mu^{\rm pc}$, are therefore irreducible representations of this symmetry group. For one dimensional representations, the PC basis vectors transform in the same representation such that the overall steady state is invariant. This implies the existence of selection rules for the overlaps $\llangle O \vert \rho_\mu^{\rm pc}\rrangle$ since only the part of $O$ that transforms like the PC basis vector, can give a finite overlap. This means that the symmetry properties of the principal components specify which changes in observables they correspond to. 
\begin{figure}
\includegraphics[width=1\linewidth]{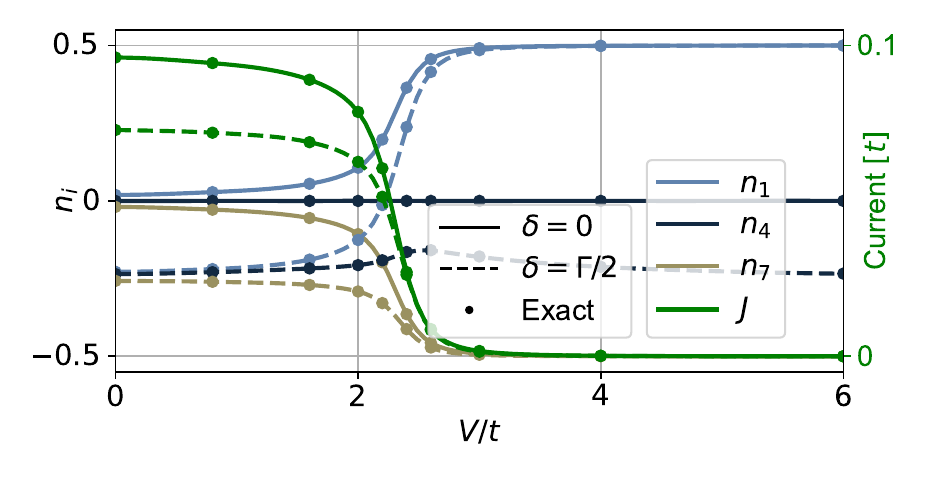}
    \caption{Occupation and current for the Fermi-Hubbard Model with $L=7$ sites. The solid and dotted lines show the observables computed from the reduced basis of dimension $d_{\rm rb}=60$ covering the parameter range $\xi =(V,\delta) \in [0,6t]\times [-\Gamma,\Gamma]$. The points show the observables computed from the exact solution.} 
    \label{fig:fermichainobs}
    
\end{figure}

\textit{Fermi-Hubbard Chain.} As a first example, we consider spinless fermions on a chain of length $L$ with Hamiltonian
\begin{align}
    H &= -t\sum_{j=1}^{L-1} \left( c_{j+1}^\dag c_j + c_j^\dag c_{j+1} \right) 
    + V\sum_{j=1}^{L-1}n_j n_{j+1} ,
\end{align}
where $c_j$ and $c_j^\dag$ are the fermionic creation and annihilation operators and $ n_j = c_j^\dag c_j-1/2$ are the particle-hole symmetrized number operators. At the endpoints the chain is coupled to two leads with chemical potentials $\mu_L$ and $\mu_R$, respectively. In the limit of large bias $\mu_L=-\mu_R \to \infty$ this can be described by a LME with two jump operators $L_1 = \sqrt{\Gamma_l}c_1^\dag$ and $L_2 = \sqrt{\Gamma_r}c_L$~\cite{oliveira2024voltagedrivenbreakdownelectronicorder}. Using a Jordan-Wigner transformation, this model maps to a driven XZZ chain which has an exact solution in the symmetric case $\Gamma_l=\Gamma_r$~\cite{prosen2011sep}. In the thermodynamic limit, $L\to \infty$, there is a phase transition between a conducting phase at $V<2t$ and an insulating phase at $V>2t$. 

We compute the reduced basis for the Liouvillian, $\hat{\mathcal L}(V,\delta)$, in the parameter range $V\in [0,6t]$ and $\delta\in [-\Gamma,\Gamma]$, where $\delta=(\Gamma_l -\Gamma_r)/2$ and $\Gamma=(\Gamma_l+\Gamma_r)/2=0.2t$ \footnote{all numerical calculations are performed using the QuTiP package~\cite{Johansson_2013}}. Fig.~\ref{fig:fermichainobs} shows the steady-state particle current, $J_j = (c_{j+1}^\dag c_j - c_{j}^\dag c_{j+1})/(2i)$ and occupations computed in the reduced-basis subspace. A comparison with the exact solutions shows that the reduced basis of dimension $d_{\rm rb}=60$ correctly captures the observables throughout the entire parameter range. 

\begin{figure}
\includegraphics[width=\linewidth]{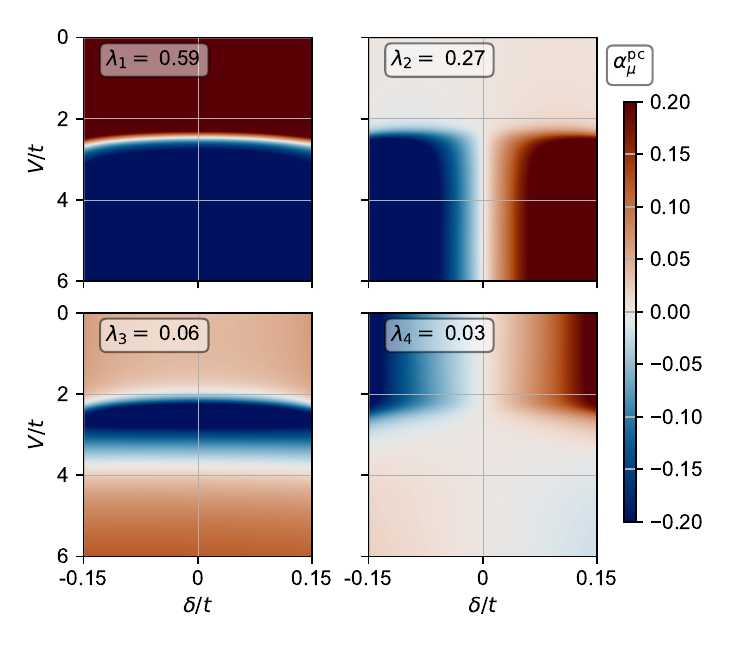}
\centering
\begin{tabular}{|c|c|c|c|c|} \hline 
&  $\chi$ & $J_j$& $n_j+n_{L-j+1}$&$n_j-n_{L-j+1}$\\ \hline 
PC1& $+$ & $\checked$& $\cross$&$\checked$\\ \hline 
PC2&$-$ & $\cross$& $\checked$&$\cross$\\ \hline 
PC3&$+$ & $\checked$& $\cross$&$\checked$\\ \hline 
PC4&$-$& $\cross$& $\checked$&$\cross$\\ \hline\end{tabular}   
\caption{Principal components, $\alpha^{\mathrm{pc}}_\mu(\xi)$, for the Fermi-Hubbard model with $L=7$ sites in the parameter range $\xi=(\delta,u)\in [-0.15t,0.15t]\times [0,3t]$ and $\mu=1,2,3,4$ with the corresponding explained variance ratios $\lambda_\mu$. The table lists the characters, $\chi$, for the PCs under the combined particle-hole and inversion transformation, together with selection rules for the overlaps, $\llangle O \vert \rho_\mu^{\rm pc}\rrangle$ for $O = J_j, n_j\pm n_{L-j+1}$.}
\label{fig:fermichain_pcmaps}
\end{figure}
The Liouvillian is invariant under a combination of particle-hole symmetry, inversion, and flipping the sign of $\delta$,
\begin{equation}
    \hat{\mathcal L}(V,\delta) = \hat{\mathcal{U}}^\dag \hat{\mathcal L}(V,-\delta) \mathcal U ,
\end{equation}
where $\hat{\mathcal U} = U\otimes U^*$ is a unitary in the Liouville space and $U$ implements the combined particle-hole and inversion transformation, $U^\dag c_j U=i(-1)^j c_{L-j+1}^\dag$ (cf. SM for a definition of $U$). 
The four PCs with the highest explained variance ratios are plotted in Fig.~\ref{fig:fermichain_pcmaps}, with the dominant ($\lambda_{1}=0.6$) component exhibiting a marked precursor to the phase transition in the thermodynamic limit. As expected, the PCs are observed to be even or odd in $\delta$, i.e. $\alpha(V,-\delta) = \chi\alpha(V,\delta)$ with characters $\chi=\pm 1$. This implies selection rules for the overlaps, $\llangle O\vert \rho_\mu^{\rm pc}\rrangle$, depending on how the observables transform under $O\mapsto U^\dag O U$. In particular, since $U^\dag J_j U = J_{L-j+1}$ and since $\langle J_j\rangle = J$ is independent of the site $j$ by charge conservation, only even PCs contribute to a change in the current according to Eq.~\eqref{eq:pctoobs}. Likewise, the inversion symmetrized (anti-symmetrized) operators $n_j + n_{L-j+1}$ ($n_j - n_{L-j+1}$) vanish for the even (odd) PCs. These symmetry properties are summarized in Fig.~\ref{fig:fermichain_pcmaps}. 

The odd PCs, $\alpha_2^{\rm pc}$ and $\alpha_4^{\rm pc}$, represent the response in the steady state to the coupling asymmetry $\delta$. As seen in Fig.~\ref{fig:fermichain_pcmaps}, they gain a finite value in one of the two phases, indicating that $\rho_2^{\rm pc}$ and $\rho_4^{\rm pc}$ measure the asymmetry in respectively the insulating and in the conducting phase. Physically this corresponds to the fact that in the insulating phase, only the central site remains susceptible to the asymmetry whereas the remaining sites are either occupied or empty. This is in contrast to the conducting phase where the system responds to the asymmetry by an overall reduction in the occupation, as seen in Fig.~\ref{fig:fermichainobs}. In the End Matter, we contrast these results to those where $L=8$, for which the explained variances exhibit qualitative differences.

\begin{figure}
    \centering
    \includegraphics[width=\linewidth]{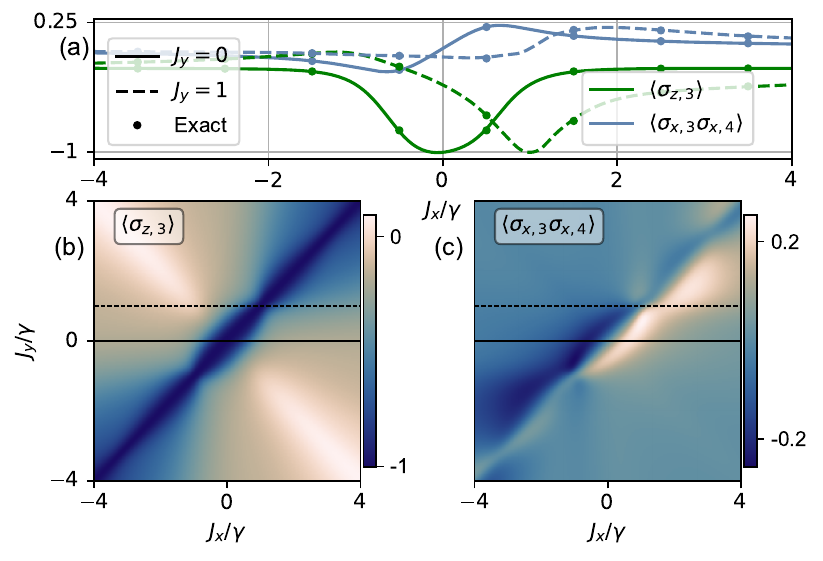}
    \caption{\textbf{(a)} $\langle \sigma_{z,3} \rangle_{\rho(\xi)}$ and $\langle \sigma_{x,3}\sigma_{x,4}\rangle_{\rho(\xi)}$ at $J_z=\gamma$ for $L=7$ sites and $J_y=0$ (solid) $J_y=1$ (dotted) computed from the reduced basis ($d_{\rm rb}=100$) covering the parameter range $\Xi=[-4\gamma,4\gamma]\times [-4\gamma,4\gamma]$ and comparison with the exact solutions (points). \textbf{(b, c)} Density plots showing respectively $\langle\sigma_{z,3}\rangle_{\rho(\xi)}$ and $\langle \sigma_{x,3}\sigma_{x,4}\rangle_{\rho(\xi)}$ in the entire parameter range.}
    \label{fig:xyz_observables}
\end{figure}

\begin{figure}[]
    \centering
    \includegraphics[width=\linewidth]{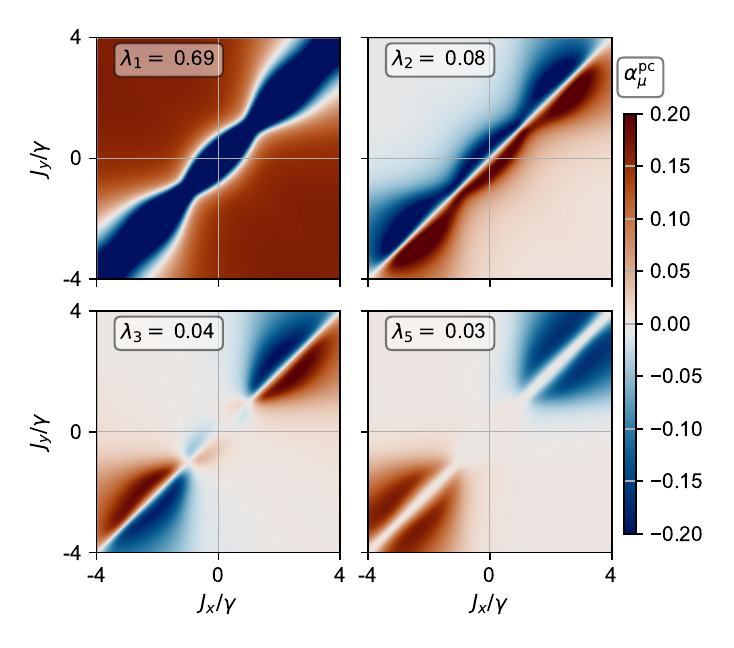}

    \begin{tabular}{|c|c|c|c|c|c|}
        \hline
         & $\chi(g_3)$&$\chi(g_4)$& $\sigma_{z,j}$& $\sigma_{x,i}\sigma_{x,j} + \sigma_{y,i}\sigma_{y,j}$&$\sigma_{x,i}\sigma_{x,j} - \sigma_{y,i}\sigma_{y,j}$\\
        \hline
        $\alpha^{\rm pc}_1$& $+$&$+$& $\checked$& $\Pi_{i+j}$& $\cross$\\ \hline 
        $\alpha^{\rm pc}_2$& $-$&$+$& $\cross$& $\cross$& $\Pi_{i+j}$\\
        \hline
 $\alpha^{\rm pc}_3$& $-$&$-$& $\cross$& $\cross$& $\Pi_{i+j+1}$\\ \hline 
 $\alpha^{\rm pc}_4$& $+$&$+$& $\checked$& $\Pi_{i+j}$& $\cross$\\\hline
 $\alpha^{\rm pc}_5$& $+$& $-$& $\cross$& $\Pi_{i+j+1}$&$\cross$\\\hline
    \end{tabular}
    \caption{PCs, $\alpha_{\mu}^{\rm pc}$, for $\mu=1,2,3,5$ along with their explained variance ratios $\lambda_\mu$ for the XYZ model with $\xi=(J_x,J_y)\in [-4\gamma,4\gamma]\times[-4\gamma,4\gamma]$ and $J_z=\gamma$. The table indicates the symmetry of the PCs under $g_3$ and $g_4$ along with the selection rules for overlaps $\llangle O\vert \rho_\mu^{\rm pc}\rrangle$. The entry $\cross$ indicates a vanishing, and $\checked$ a finite overlaps and $\Pi_k = \checked$ ($\Pi_k = \cross$) for even (odd) $k$.}
    \label{fig:xyz_pcmaps}
\end{figure}

\textit{XYZ-model}. As a second example, we consider the one-dimensional dissipative XYZ model~\cite{lee_unconventional_2013,li_steady-state_2021,chan_limit-cycle_2015}, with Hamiltonian
\begin{align}
H=\!\sum_{j=1}^{L-1}\!\left(J_x \sigma_{x,j}\sigma_{x,j+1} +J_y \sigma_{y,j}\sigma_{y,j+1}+J_y \sigma_{y,j}\sigma_{y,j+1} \right)
\end{align}
and dissipation on each site described by the jump operators $L_j =\sqrt{\gamma}(\sigma_{x,j}-i\sigma_{y,j})/2$. We construct the reduced basis ($d_{\rm rb} = 100$) for the parameters $\xi = (J_x,J_y)$ in the range $\Xi = [-4\gamma,4\gamma]\times [-4\gamma,4\gamma]$ with $J_z=\gamma$, where the model is known to exhibit a rich mean-field phase diagram ~\cite{lee_unconventional_2013}. In Fig.~\ref{fig:xyz_observables} we show the expectation values of $\sigma_{z,j}$ and $\sigma_{x,i}\sigma_{x,j}$ and compare with the exact solutions at selected points, showing that the reduced basis correctly captures the complicated parameter dependencies of the observables (See Fig.~1 in the Supplemental Material for a comparison with $d_{\rm rb}=20$ and $d_{\rm rb}=50$). 
This model has four symmetries, two of which leave $\xi$ invariant while the other two, denoted $g_3$ and $g_4$, transform $\xi$ as $ (J_x,J_y) \stackrel{g_3}{\longmapsto} (J_y,J_x)$ and $ (J_x,J_y) \stackrel{g_4}{\longmapsto} (-J_x,-J_y)$ (cf. SM). As before, the PCs form irreducible representations of the symmetry group with their respective characters indicated in the first two columns of the table in Fig.~\ref{fig:xyz_pcmaps} for the 5 PCs with highest explained variance ratios. By inspecting the commutation relations between the observables and the representations of the symmetry transformations on the Hilbert space (shown in detail in SM), one can find the selection rules for any observables of interest as is shown in the last three columns of the table in Fig.~\ref{fig:xyz_pcmaps}. 

Fig.~\ref{fig:xyz_pcmaps} shows that the predominant state variance in this parameter range occurs along $\rho^{\rm pc}_{1}$. That is, the state variance projected onto $\rho^{\rm pc}_{1}$ (cf. Eq.~\ref{eq:cmudef}) accounts for $69\%$ of the total state variance. The similarity between $\alpha_1^{\rm pc}(\xi)=\expval{\rho^{\rm pc}_{1}}_{\rho(\xi)}-\expval{\rho^{\rm pc}_{1}}_{\rho_{0}}$ and $\expval {\sigma_{z,3}}_{\rho(\xi)}$ (apparent from Figs.~\ref{fig:xyz_observables}(b) 
and~\ref{fig:xyz_pcmaps}) combined with the fact that $\sigma_{z,3}$ has the same symmetry properties as $\rho_1^{\rm pc}$ indicates a large overlap, $\llangle \sigma_{z,3}\vert \rho_1^{\rm pc}\rrangle$. The PC basis vectors with smaller explained variances that have the same symmetry properties as $\rho_1^{\rm pc}$ also have a finite overlap with $\sigma_{z,3}$ accounting for the differences between $\alpha_1^{\rm pc}$ and $\expval {\sigma_{z,3}}_{\rho(\xi)}$. 

Note that the PCs are irreducible representations of the symmetry group only up to statistical noise coming from either the sampling in parameter space or the small error in the reduced basis. This tends to get worse as the explained variance gets smaller, since the smaller explained variances are closer to accidental degeneracies, making them more susceptible to weak symmetry breakings.

\textit{Discussion.} 
The success of the RBM in rapidly reaching a high accuracy hinges on the existence of patches in parameter space with a high linear dependence of the steady states. As a result, one would generically expect a reduced performance of the RBM with increasing system size since this implies smaller overlaps of the solutions at different parameter points, which in turn requires more basis vectors to achieve the desired accuracy. This question was investigated for closed systems in  Ref.~\onlinecite{Brehmer2023Apr} by considering the scaling of $d_{\rm rb}$ with system size and it would be an interesting future direction to perform a similar study for open 1D systems using matrix product states~\cite{weimer_simulation_2021, somoza_dissipation-assisted_2019}. 

In Ref.~\onlinecite{minganti_spectral_2018}, a dissipative quantum phase transition of order $m$ ($m=1,2$) was defined as a point in parameter space, $\xi_c$, at which $\lim_{\xi \to \xi_c}|\partial^{m}_{\xi}\lim_{L\mapsto \infty}\expval{O}_{\rho_{\mathrm{\rm ss}(\xi,L)}}|=\infty$, where $L$ is a parameter that controls the thermodynamic limit (e.g. the length of a 1D chain) and $O$ is a $\xi$-independent observable. For problems, which are amenable to the RBM, i.e. where $d_{\rm rb}$ scales favorably with system size, the PC approach could provide an efficient means of identifying the order parameters of dissipative phase transitions as the PC basis vectors, $O=\rho_\mu^{\rm pc}$, corresponding to $\alpha_\mu^{\rm pc}(\xi)$ with a precursor to a discontinuity (i.e. $\mu=1$ in Fig.~\ref{fig:fermichain_pcmaps}). We note that this approach would be entirely 'data-driven' in the sense that it does not require any a priori knowledge of the order parameters characterizing the steady-state phases of the system. This would extend the repertoire of recent interesting ideas for using unsupervised machine learning for detecting phase transition in the context of equilibrium~\cite{wang_discovering_2016, van_nieuwenburg_learning_2017, zang_machine_2024, wetzel_unsupervised_2017} and non-equillibrium~\cite{bohrdt_analyzing_2021, bhakuni2024} quantum systems. 





\begin{acknowledgments}
 
We thank Alexandra Haslund-Gourley for useful discussions and suggestions. H.~C. acknow\-ledges support from the Novo Nordisk Foundation grant NNF20OC0060019. V.V.B. was supported by a grant of the Romanian Ministry of Education and Research, Project No. 760122/31.07.2023 within PNRR-III-C9-2022-I9.
\end{acknowledgments}

\bibliography{bibliography}


\subsection{End Matter}

\begin{figure}[h!]
    \centering
    \includegraphics[width=\linewidth]{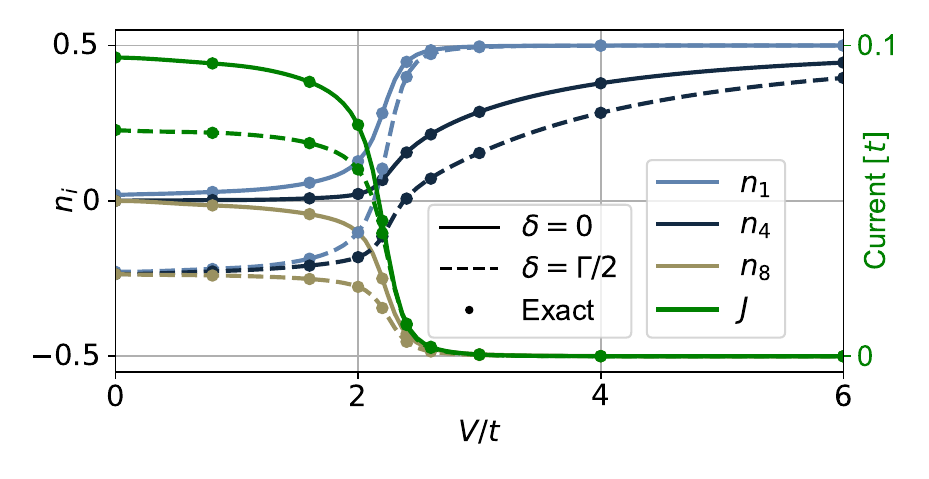}
    \caption{Occupation and current for the Fermi-Hubbard Model with $L=8$ sites computed with the reduced basis ($d_{\rm rb}=64$) in the same parameter range as in Fig.~\ref{fig:fermichainobs}.}
    \label{fig:fermichain_obs_8site}
\end{figure}

In the following, we compare the Fermi-Hubbard model with $L=8$ sites (Fig.~\ref{fig:fermichain_obs_8site}) to the $L=7$ results in the main text (Fig.~\ref{fig:fermichainobs}), with the main qualitative difference being the absence of a central site when $L$ is even. 

\begin{figure}[h!]
    \centering
    \includegraphics[width=\linewidth]{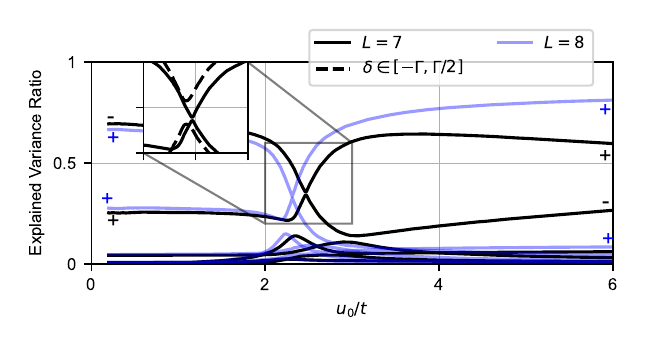}
    \caption{Explained variance ratios (normalized spectrum of $C^{\Xi_{u_0}}$) as the parameter range $\Xi_{u_0}=[0,u_0]\times [-\Gamma,\Gamma]$ is changed from $u_0=0.1t$ to $u_0=6t$ for the Fermi-Hubbard model with chain length $L=7$ (black) and $L=8$ (blue). The two largest explained variance ratios are marked with a symbol indicating whether their corresponding PCs are symmetric ($+$) or anti-symmetric ($-$) under the particle-hole transformation. The dotted line in the inset is calculated for the parameter range $\Xi_{u_0}=[0,6t]\times [-\Gamma,\Gamma/2]$ and exhibits an avoided level crossing since the parameter range now explicitly breaks the particle-hole symmetry transformation $\delta\mapsto -\delta$.}
    \label{fig:fermichain_evrspectrum}
\end{figure}

In Fig.~\ref{fig:fermichain_evrspectrum}, we show the spectrum of the explained variance ratios, $\lambda_\mu$, corresponding to the covariance matrix $C^{\Xi_{u_0}}$ as the parameter range $\Xi_{u_0}=[0,u_0]\times [-\Gamma,\Gamma]$ is varied from $u_0=0.1t$ to $u_0=6t$. At $u_0\leq 2t$, there is a dominant symmetric PC ($\lambda_1\approx 0.7$) and a subdominant anti-symmetric PC ($\lambda_2\approx 0.25$) for both even and odd $L$ meaning that there is both a particle-hole symmetric and anti-symmetric response to changes in $\delta$. At $u_0\approx 2t$ the spectrum features a level crossing between symmetric and anti-symmetric branches which signals a precursor to the phase transition in the thermodynamic limit~\cite{prosen2011sep}. Finally, for $u_0>2t$, the asymptotic behavior depends on the parity of $L$. For even $L$, the symmetric PC dominates completely, indicating that all the sites are locked to be fully occupied on the left half of the chain while being empty on the right half. For odd $L$, the remaining subdominant asymmetric PC with $\lambda_2\approx 0.25$ indicates that there is a remaining susceptibility to $\delta$ in the occupation of the central site. This provides an illustration of how the explained variance spectrum can provide an observable-independent analysis of the response to symmetry breaking perturbations.

\end{document}


\title{Reduced Basis Method for Driven-Dissipative Quantum Systems: Supplementary Material}

\author{Hans Christiansen$^{1}$}
\author{Virgil V. Baran$^{1,2,3\,}$}
\author{Jens Paaske$^{1}$}
\affiliation{$^{1}$Center for Quantum Devices, Niels Bohr Institute, University of Copenhagen, 2100 Copenhagen, Denmark}
\affiliation{$^{2}$Faculty of Physics, University of Bucharest, 405 Atomi\c stilor, RO-077125, Bucharest-M\u agurele, Romania}
\affiliation{$^{3}$"Horia Hulubei" National Institute of Physics and Nuclear Engineering, 30 Reactorului, RO-077125, Bucharest-M\u agurele, Romania}

\date{May 8, 2025}

\begin{abstract}
    This supplemental material presents some details of the derivations of the results presented in the main text.
\end{abstract}

\maketitle

\tableofcontents

\section{Lindblad Master Equation}\label{prelim}

The Lindblad master equation (LME) describing the density matrix of an open quantum system of dimension $d_{\mathcal{H}}$ can be written as 

\begin{align}
    \frac{d\vert\rho\rrangle}{dt} = \mathcal L \vert \rho \rrangle 
\end{align}
where the \textit{Liouvillian}, $\mathcal L$, is a linear super operator acting on the vectorized density matrix $\vert \rho \rrangle$ of dimension $d_{\mathcal L} = d_{\mathcal H}^2$ and is written as
\begin{align}\label{eq:liouhermdissipativejump}
    \mathcal L = \mathcal L_H + \mathcal L_D + \mathcal L_J ,\
\end{align}
where $\mathcal L_H$, $\mathcal L_D$, and $\mathcal L_J$ describes the coherent, dissipative and jump terms, respectively and are defined as \cite{minganti_spectral_2018}
\begin{align}\label{eq:liouvillestructure}
    \mathcal L_H &= -i( H \otimes \mathds 1 - \mathds 1 \otimes H^*),\ \notag \\
    \mathcal L_D &= -\sum_j \qty(L_j^\dag L_j \otimes \mathds 1 + \mathds 1 \otimes \qty(L_j^\dag L_j)^*) ,\ \notag \\
    \mathcal L_J &= \sum_j 2 L_j \otimes L_j^* .\
\end{align}
In the following, we assume that $\mathcal L $ is diagonalizable, so we can write it in terms of its right and left eigenvectors $\vert \rho_n \rrangle$ and $\vert l_n \rrangle$ as \footnote{throughout this work, we will assume that the Liouvillian is diagonalizable~\cite{horn2012matrix}}
\begin{align}
    \mathcal L = \sum_n \lambda_n \vert \rho_n \rrangle \llangle l_n \vert .\
\end{align}
The left and right eigenvectors can be chosen to satisfy the bi-orthonormality condition $\llangle l_n \vert r_m \rrangle =\delta_{nm}$. The time evolution of an initial state $\vert \rho(t=0)\rrangle = \sum_n c_n \vert \rho_{n}\rrangle$ is then given by
\begin{align}\label{eq:vectorizedtimeevolution}
    \vert \rho(t) \rrangle = \sum_n c_n\mathrm{e}^{\lambda_n t} \vert \rho_n \rrangle .\
\end{align}
The normalization of the reduced density matrix as $\Tr\{\rho(t)\}= 1$ at all times $t$, implies that $\Tr\{\mathcal L \rho\}=0 = \llangle \mathds 1 \vert \mathcal L \vert \rho \rrangle$. Here $\vert \mathds 1 \rrangle$ is the vectorized identity operator. Since this is true for all $\rho$ this means that $\llangle \mathds 1 \vert \mathcal L = 0$, implying that $\llangle \mathds 1 \vert$ is the left eigenvector of $\mathcal L$ with eigenvalue $\lambda_0=0$. The corresponding right eigenvector is the steady state. 
The remaining eigenvalues must all have non-positive real parts to ensure that the solution converges to the steady states for long times. This property can also be directly related to the algebraic structure of Eq.~\eqref{eq:liouvillestructure} \cite{zhang_direct_2025}. 

A Liouvillian is said to have a \textit{weak} symmetry if a unitary operator $\hat{\mathcal U}$ commutes with the Liouvillian, $[\hat {\mathcal L},\hat{\mathcal U}] = 0$, meaning that $\hat{\mathcal L}$ can be decomposed as 
\begin{align}
    \hat{\mathcal L} = \bigoplus_n \hat{\mathcal L}_{u_n} ,\
\end{align}
where $\hat{\mathcal L}_{u_n}$ are the symmetry blocks corresponding to the eigenvalues $u_n$ of $\hat{\mathcal U}$~\cite{buca_note_2012}. When the weak symmetry admits the decomposition $\hat{\mathcal U}=U\otimes U^*$, the steady state vector must always belong to the symmetry sector with $u_0=1$ since we can write 
\begin{align}
    u = \llangle \mathds 1 \vert \hat{\mathcal U} \vert \rho_{ss} \rrangle =  \Tr{U\rho_{ss}U^\dag} = 1,\
\end{align}
where in the last equality, we use the cyclic property of the trace along with the normalization condition of the steady state. From this we conclude that the steady state always transform trivially under weak symmetries of the Liouvillian.

\section{Principal Component Basis}

In this section, we will describe the transformation to the principal component (PC) basis. Suppose we have a reduced basis such that the steady state of the Liouvillian, $\mathcal L(\xi)$, at each parameter point $\xi$ can be written as 
\begin{align}\label{eq:rhorbrep}
    \vert \rho(\xi) \rrangle =  \alpha_\mu(\xi)\vert \rho_\mu \rrangle = \vert \rho_0 \rrangle +  \tilde{\alpha}_\mu(\xi) \vert \tilde{\rho}_\mu \rrangle ,\
\end{align}
where in the following, all Greek indices run over the reduced basis dimension $\mu=1,..d_{\rm rb}$, and we use the convention that repeated indices are summed over. In the second equality, we use the shifted coefficients $\alpha_\mu \mapsto \tilde{\alpha}_\mu=\alpha_\mu - \langle \alpha_\mu\rangle_\xi$, where $\expval{\cdot}_{\Xi}$ denotes averaging over the parameter space, and $\vert \rho_0 \rrangle = \sum_{\mu}\expval{\alpha_\mu}_\Xi\vert \rho_\mu\rrangle$ represents the mean density matrix. We now form the covariance matrix of the $\tilde{\alpha}_\mu(\xi)$ as 
\begin{align}
    C^\Xi_{\mu\nu} = \left\langle\tilde{\alpha}_\mu\tilde{\alpha}_\nu \right\rangle_{\Xi} .\
\end{align}
We diagonalize this positive symmetric matrix in terms of the orthogonal matrix $W$ as
\begin{align}
    W^T_{\eta\mu}\langle \tilde{\alpha}_\mu \tilde{\alpha}_\nu \rangle_{\Xi}W_{\nu \delta} = c_{\eta} \delta_{\eta\delta}, 
\end{align}
where the real positive eigenvalues, $c_\mu$, are arranged in descending order such that $c_1\geq c_2\geq ...c_{d_\mathrm{rb}}$. Let $\alpha_{\nu}^{\rm pc}(\xi)=\tilde{\alpha}_\mu(\xi)W_{\mu\nu}$ with the corresponding PC basis vectors $\vert \rho_\nu^{\mathrm{pc}}\rrangle = W^T_{\nu\mu}\vert{\rho_{\mu}}\rrangle$, so Eq.~\eqref{eq:rhorbrep} can be written as 
\begin{align}\label{eq:rhorbrep}
    \vert \rho(\xi) \rrangle = \vert \rho_0 \rrangle +\tilde{\alpha}_\mu^{\rm pc}(\xi) \vert \tilde{\rho}_\mu^{\rm pc} \rrangle .\
\end{align}
The expectation value of an observable, $O$, is then given by 
\begin{align}\label{eq:alphatoobs}
    \expval{O}_{\rho(\xi)} &= \llangle O \vert \rho_0\rrangle + \alpha_\mu^{\mathrm{pc}}(\xi)\llangle O \vert \rho_\mu^{\mathrm{pc}}\rrangle .\
\end{align}
We consider transformations $g$ that act on the Liouvillian as 
\begin{align}\label{eq:gensymtransformation}
    g[\hat{\mathcal L}(\xi)] =  \hat{\mathcal U}^\dag(g)\hat{\mathcal L}(u(g^{-1})\xi)\hat{\mathcal{U}}(g) ,\
\end{align}
where $\mathcal U(g)=U(g)\otimes U^*(g)$ is the representation in the Liouville space and $u$, the representation in parameter space. We consider the group of symmetry operators $g$ that leave the parameter space $\Xi$ invariant and the Liouvillian invariant according to the transformation in Eq.~\eqref{eq:gensymtransformation}. From this we can write
\begin{align}
    0 = \hat{\mathcal L}(\xi)\vert \rho(\xi)\rrangle = \hat{\mathcal{U}}^\dag(g) \hat{\mathcal L}(u(g^{-1})\xi)\hat{\mathcal{U}}(g)\vert \rho(\xi) \rrangle ,\
\end{align}
and conclude that $\hat{\mathcal{U}}(g)\vert \rho(\xi)\rrangle=\vert U(g)\rho(\xi)U^\dag(g)\rrangle$ is equal to the steadystate at $u(g^{-1})\xi$, 
\begin{align}\label{eq:symtransdm}
    \rho(u(g^{-1})\xi) = U(g)\rho(\xi)U^\dag(g) ,\
\end{align}
where we used that $U(g^{-1})=U^{-1}(g) = U^\dag (g)$. We can also write Eq.~\eqref{eq:symtransdm} in vectorized notation as 
\begin{align}\label{eq:rhosymtransform}
    \vert \rho(u(g)\xi)\rrangle = \hat{\mathcal U}^\dag(g) \vert \rho(\xi)\rrangle .
\end{align}
Now we write Eq.~\eqref{eq:rhosymtransform} in terms of the reduced basis vectors as 
\begin{align}\label{eq:symrepvectorized}
    \tilde{\alpha}_\mu(u(g)\xi)\vert \rho_\mu \rrangle = \tilde{\alpha}_\mu(\xi)\hat{\mathcal U}^\dag(g)\vert \rho_\mu\rrangle .\
\end{align}
if the reduced basis accurately captures the symmetries of the model then $\hat{\mathcal U}^\dag (g)\vert \rho_\mu\rrangle$ must stay within the reduced-basis subspace, i.e. we can write Eq.~\eqref{eq:symrepvectorized} as 
\begin{align}\label{eq:rhosymrelation}
    \tilde{\alpha}_\mu(\xi) \vert \rho_\mu \rrangle = \tilde{\alpha}_\nu(u^{-1}(g)\xi)V_{\nu\mu}^{-1}(g)\vert \rho_{\mu}\rrangle ,\
\end{align}
where $V_{\mu\nu}^{-1}(g) = \llangle \rho_\mu \vert \hat{\mathcal{U}}^\dag (g) \vert \rho_\nu\rrangle = V_{\mu\nu}^T$ is the representation in the reduced basis. For this equation to hold, the $\tilde{\alpha}_\mu(\xi)$ vectors must transform as $\tilde{\alpha}_\nu(u^{-1}(g)\xi)=\tilde{\alpha}_\lambda(\xi)V_{\lambda\nu}(g)$, from which we conclude that 
\begin{align}\label{eq:covariancesymmetry}
    V_{\lambda \mu}C^{\Xi}_{\mu\nu} V^T_{\nu\eta} &= \langle V_{\lambda\mu}\tilde{\alpha}_\mu(\xi)V_{\eta\nu}\tilde{\alpha}_\nu(\xi)\rangle_\Xi 
    \notag \\ &= \langle \tilde{\alpha}_\lambda(u(g)\xi)\tilde{\alpha}_\eta(u(g)\xi)\rangle_\Xi 
     \notag \\ &= \langle \tilde{\alpha}_\lambda(\xi)\tilde{\alpha}_\eta(\xi)\rangle_{u(g)\Xi}
    = C_{\lambda\eta}^\Xi ,\    
\end{align}
where in the last equality, we used that the parameter space is invariant under $u(g)$. From Eq.~\eqref{eq:covariancesymmetry}, we learn that if the Liouvillian is invariant under a symmetry group $G$ according with \eqref{eq:gensymtransformation}, this implies that representation of $g$, $V(g)$, exists such that 
\begin{align}
    [V(g),C^\Xi]=0 ,\
\end{align}
for all $g$. In particular this means that the eigenvectors of $C^\Xi$, $\alpha^{\rm pc}_\mu$, transform as irreducible representations of $G$. 
Now let us restrict to 1-dimensional irreps such that the principal components transform as $\alpha_\mu^{\mathrm{pc}}(O(g)\xi)=\chi_\mu(g)\alpha_\mu^{\mathrm pc}(\xi)$, where $\chi_\mu(g)$ is the character of the transformation $g$ in the irreducible representation corresponding to the $\mu$'th eigenvalue. To satisfy the symmetry relation Eq.~\eqref{eq:rhosymrelation} we then conclude that the PC basis vectors transform in the same representation. From this we see that the overlap $\llangle O \vert \rho_\mu^{\rm pc}\rrangle$ is only non-zero if $O$ and $\rho^{\mathrm{PC}}_\mu$ belong to the same representation.

\section{Fermi-Hubbard Chain}~\label{sec:fermihubbardsec}

The Fermi-Hubbard chain is described by the Hamiltonian 
\begin{align}
    H &= -t\sum_{j=1}^{L-1} \left( c_{j+1}^\dag c_j + c_j^\dag c_{j+1} \right) 
    + V\sum_{j=1}^{L-1}n_j n_{j+1} ,\
\end{align}
and jump operators $L_1 = \sqrt{\Gamma_l}c_1^\dag$ and $L_2 = \sqrt{\Gamma_r}c_L$ where we use $n_j = c_j^\dag c_j -\frac{1}{2}$ as in the main text. Defining $2\Gamma \equiv \Gamma_l +\Gamma_r $ and $2\delta = \Gamma_l -\Gamma_r$ the dissipative and jump parts of the Liouvillian become 
\begin{align}
    \hat{\mathcal L}_D &= -\Gamma \left( (1-n_1+n_L) \otimes \mathds  1 + \mathds 1 \otimes  (1-n_1+n_L)  \right) \notag \\
    & \phantom{=\:}+\delta \left( (n_1+n_L) \otimes \mathds  1 + \mathds 1 \otimes  (n_1+n_L)  \right),\
\end{align}
and 
\begin{align}
    \hat {\mathcal L}_J &= 2\Gamma (c_1^\dag \otimes c_1^\dag + c_L\otimes c_L ) + 2\delta (c_1^\dag \otimes c_1^\dag - c_L\otimes c_L ) .\
\end{align}
This model has a particle-hole type symmetry of the form of Eq.~\eqref{eq:gensymtransformation} with $\xi =(u,\delta) \mapsto (u,-\delta)$ and unitary $\mathcal U = U\otimes U^*$, where $U$ is an operator in the Hamiltonian Hilbert space implementing a particle-hole and inversion symmetry, $U=U_{\rm inv}U_{\rm ph}$. The particle-hole transformation is given by
\begin{equation}
    U_{\rm ph} = \prod_{j=1}^{L}(c_j - (-1)^{j}c_j^\dag),
\end{equation}
transforming the fermionic operators as 
\begin{align}
U_{\rm ph}^\dag c_j U_{\rm ph} &= (-1)^{L-1}(c_j^\dag-(-1)^{j}c_j) c_j (c_j - (-1)^{j}c_j^\dag) \notag \\
    &=(-1)^{j+L}c_j^\dag.
\end{align}
Defining an anti-Hermitian operator as
\begin{equation}
    S=i\frac{\pi}{2}\sum_{j=1}^{\lfloor L/2\rfloor}(c_{j}^\dag c_{L-j+1} + c_{L-j+1}^\dag c_j),
\end{equation}
the inversion transformation is given by
\begin{align}
\label{eq:invtransLodd}
U_{\rm inv} &=
\begin{cases}
\mathrm{e}^{-S}\mathrm{e}^{-i\frac{\pi}{2}c_{(L+1)/2}^\dag c_{(L+1)/2}}, & \text{if } L \text{ odd}, \\
\mathrm{e}^{-S}, & \text{if } L \text{ even}.
\end{cases}
\end{align}
To see this, consider the unitary transformation $\tilde{A}(\eta) \equiv {\rm e}^{\eta S}A {\rm e}^{-\eta S}$. For $j\neq (L+1)/2$ we find the operator equation \cite{bruus_2004}
\begin{align}
    \partial_{\eta}\mqty( \tilde{c}_{j} \\ \tilde{c}_{L-j+1})(\eta) &= \mqty([{S}(\eta),\tilde{c}_{j}(\eta)] \\ [S(\eta),\tilde{c}_{L-j+1}(\eta)])\notag\\
    &= - i\frac{\pi}{2}\mqty(\tilde{c}_{L-j+1} \\ \tilde{c}_{j})(\eta)  \notag \\
    & = -i\frac{\pi}{2}\sigma_{x}\mqty(\tilde{c}_{j} \\ \tilde{c}_{L-j+1})(\eta) .
\end{align}
Solving this and evaluating at $\eta=1$, we then find the transformation
\begin{equation}
    U_{\rm inv}^\dag c_j U_{\rm inv} = -ic_{L-j+1} .
\end{equation}
For $L$ odd, the extra factor in Eq.~\eqref{eq:invtransLodd} means that the middle site transforms as $U^\dag_{\rm inv}c_{(L+1)/2}U_{\rm inv}=-ic_{(L+1)/2}$. Under the combined symmetries, the operators transform as 
\begin{align}
    U^\dag c_j U &= i(-1)^j c^\dag_{L-j+1},\notag \\
    U^\dag c_j^\dag U &= -i(-1)^j c_{L-j+1}.
\end{align}
The density operators transform as 
\begin{align}\label{eq:pcommute}
    U^\dag n_j U = - n_{L-j+1},
\end{align}
while the tunneling terms transform as 
\begin{align}\label{eq:tunnellingtransform}
    U^\dag (c_j^\dag c_{j+1} + c_{j+1}^\dag c_j ) U &= (-1)^{2j+1} (c_{L-j+1} c_{L-r}^\dag + c_{L-j} c_{L-j+1}^\dag) \notag \\
    &= c_{L-j}^\dag c_{L-j+1} + c_{L-j+1}^\dag c_{L-j} .
\end{align}
From this we see that the the full Liouvillian transforms as
\begin{align}\label{eq:liouphs}
    \hat{\mathcal U}^\dag \hat{\mathcal L}(V,\delta) \hat{\mathcal U} = \hat{\mathcal L }(V,-\delta) .\
\end{align}
as written in the main text.

\section{XYZ Model}

\begin{figure}
    \centering
    \includegraphics[width=\linewidth]{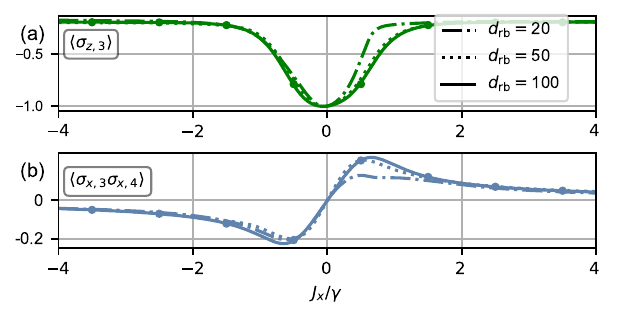}
    \caption{\textbf{(a)} $\langle \sigma_{z,3} \rangle_{\rho(\xi)}$ and \textbf{(b)} $\langle \sigma_{x,3}\sigma_{x,4}\rangle_{\rho(\xi)}$ (b) at $J_z=\gamma$ and $J_y=0$ for $L=7$ sites computed from the reduced basis $d_{\rm rb}=20$ (dot-dashed), $d_{\rm rb}=50$ (dotted) and $d_{\rm rb}=100$ (solid) covering the parameter range $\Xi=[-4\gamma,4\gamma]\times [-4\gamma,4\gamma]$, and comparison with the exact solutions (points).}
    \label{fig:xyz_dcomparison}
\end{figure}

The dissipative XYZ model consists of the Hamiltonian 
\begin{align}
    H = \sum_j \left(J_x \sigma_{x,j}\sigma_{x,j+1} + J_y \sigma_{y,j}\sigma_{y,j+1} + J_z \sigma_{z,j}\sigma_{z,j+1}\right)
\end{align}
along with the jump operators $L_j =\sqrt{\gamma}\sigma_{-,j}$, resulting in the Liouvillian written as in Eq.~\eqref{eq:liouhermdissipativejump} with 
\begin{align}
    \hat{\mathcal L}_D &= -\gamma\sum_{j=1}^L \left[\sigma_{-,j}^\dag \sigma_{-,j+1} \otimes \mathds 1 + \mathds 1 \otimes (\sigma_{-,j}^\dag \sigma_{-,j+1})^* \right] \notag \\
    &= -\frac{\gamma}{2}\sum_{j=1}^L \left[(\sigma_0 + \sigma_z)\otimes \mathds 1 + \mathds 1 \otimes (\sigma_0 +\sigma_z)\right] ,\
\end{align}
and 
\begin{align}
    \hat{\mathcal L}_J &= 2\gamma \sum_{j=1}^L \sigma_{-,i} \otimes \sigma_{-,i}  \notag \\ &=\frac{\gamma}{2}\left[\sigma_{x,j} \otimes \sigma_{x,j} - \sigma_{y,j}\otimes \sigma_{y,j}\right.  \notag \\ 
    &\phantom{=\frac{\gamma}{2}}\left.\,\,-i(\sigma_{x,j} \otimes \sigma_{y,j}+\sigma_{y,j} \otimes \sigma_{x,j})\right]
\end{align}
In Fig.~\ref{fig:xyz_dcomparison} we show the expectation values of $\sigma_{z,3}$ and $\sigma_{x,3}\sigma_{x,4}$ with different reduced basis dimensions, $d_{\rm rb}$. This model has two symmetries that leave the parameters invariant and two symmetries that also transform the parameters according with Eq.~\eqref{eq:gensymtransformation}, where all the symmetries are of the form $\mathcal U_i \equiv \mathcal U(g_i) = U_i\otimes U_i^*$. First we consider the symmetries that leave the parameters invariant. These are given by the string operator $U_1 =\prod_i \sigma_{z,i}$ and the string of two-qubit swap gates \cite{Nielsen_Chuang_2010}
\begin{align}
    U_2 = \prod_{r=1}^{\lfloor \frac{L}{2} \rfloor}\mathrm{SWAP}_{r,L-r+1} ,\
\end{align}
where $\mathrm{SWAP}_{r,r'}=\frac{1}{2}(\mathds 1 + \sum_{i\in\{x,y,z\}}\sigma_{i,r}\sigma_{i,r'})$ for $r\neq r'$ is a unitary and hermitian operator that acts as 
\begin{align}
    \mathrm{SWAP}_{r,r'} \sigma_{i,r} \mathrm{SWAP}_{r,r'} = \sigma_{i,r'} .\
\end{align}
Next, we consider the unitary $\hat{\mathcal{U}}_3 = U_3\otimes U_3^*$ with 
\begin{align}
    U_3 = \prod_{j=1}^L \frac{\sigma_{0,j}+i\sigma_{z,j}}{\sqrt{2}} .\
\end{align}
$U_3$ commutes with $\sigma_{z,j}$ while it transforms $\sigma_{y,j}$ and $\sigma_{x,j}$ as 
\begin{align}
    U_3^\dag \sigma_{x,j} U_3 = \sigma_{y,j} ,\ && U_3^\dag \sigma_{y,j} U_3 = -\sigma_{x,j} ,\
\end{align}
implying that in the second tensor factor of the Liouvillian space, the spin operators transform with the signs swapped 
\begin{align}
    U_3^T \sigma_{x,j} U_3^* = -\sigma_{y,j} ,\ && U_3^T \sigma_{y,j} U_3^* = \sigma_{x,j} .\
\end{align}
From this we see that the Liouvillian transform as 
\begin{align}\label{eq:u2transform}
    \hat{\mathcal U}_3^\dag\mathcal L(J_x,J_y,J_z) \hat{\mathcal{U}}_3 = \mathcal L(J_y,J_x,J_z)
\end{align}
Finally, under the symmetry $\hat{\mathcal U_4} = U_4 \otimes U_4^*$ with
\begin{align}
    U_4 = \prod_{j \: \mathrm{even}} \sigma_{z,j} ,\
\end{align}
the Liouvillian transforms as 
\begin{align}\label{eq:u3transform}
    \hat{\mathcal U}_4^\dag \mathcal L(J_x,J_y,J_z) \hat{\mathcal U_4} = \mathcal L(-J_x,-J_y,J_z) .\
\end{align}
To understand the selection rules of the overlaps $\llangle O\vert \rho_\mu^{\rm pc}\rrangle$ in Eq.~\eqref{eq:alphatoobs} we consider how the symmetries that act non-trivially on the parameter space transform the observables of interest. For example, the $\sigma_{z,j}$ is invariant under both $\hat{\mathcal U}_3$ and $\hat{\mathcal U}_4$ so $\llangle O\vert \rho_\mu^{\rm pc}\rrangle$ is only non-vanishing when the corresponding principal component is even under both $g_3$ and $g_4$. $O^\pm_{j,j'} \equiv \sigma_{x,j}\sigma_{x,j'}\pm \sigma_{y,j}\sigma_{y,j'}$ transforms as 

\begin{align}
    U_3^\dag O^\pm_{j,j'} U_3 &= \pm O_{j,j'}^\pm, \notag \\ U_4^\dag O_{j,j'}^\pm U_4 &= (-1)^{j-j'}O^\pm_{j,j'} .\
\end{align}

from which the selection rules written in the main text can be derived.

\section{Bound on the Reduced-Basis error}

We consider the steadystate residual defined as 
\begin{align}\label{eq:ssresdef}
    \mathrm{Res}(\xi) = \frac{\llangle \rho^{\mathrm{rb}}\vert \mathcal L^\dag \mathcal L \vert \rho^{\mathrm{rb}}\rrangle}{\llangle \rho^{\mathrm{rb}}\vert \rho^{\mathrm{rb}}\rrangle} ,\
\end{align}
where in the following we will leave the $\xi$ dependence implicit. Consider the set of normalized eigenvectors $\vert M_n\rrangle$ and real eigenvalues $M_n$ of the positive semi-definite operator $M=\mathcal L^\dag \mathcal L$. We sort $M_n$ in ascending order, $M_0 \leq M_1 \leq .. \leq M_{d}$, such that the first eigenvalue $M_0 = 0$ with the corresponding eigenvector proportional to the exact steadystate, $\vert M_0 \rrangle =  \vert \rho_{\mathrm{ex}}\rrangle/\norm{\rho_{\rm ex}}$. Let $M_1 =\Gamma>0$ be the spectral gap of $M$. We now expand $\vert \rho^{\mathrm{rb}}\rrangle $ in the orthonormal basis $\vert M_n\rrangle$ as 
\begin{align}
    \vert \rho^{\mathrm{rb}}\rrangle =  c_0 \vert \rho^{\mathrm{ex}}\rrangle + \sum_{n>0} c_n \vert M_n \rrangle \equiv c_0 \vert \rho^{\mathrm{ex}} \rrangle + \vert \delta \rho \rrangle,\
\end{align}
where $\vert \delta\rho\rrangle$ is $\vert \rho^{\mathrm{rb}}\rrangle$ projected onto the orthogonal complement of $\vert \rho^{\mathrm{ex}}\rrangle$ and $c_0 = 1-\llangle \mathds 1 \vert \delta \rho\rrangle$ to ensure that $\llangle \mathds 1 \vert \rho^{\mathrm{rb}}\rrangle =1$. The residual can now be lower bounded as
\begin{align}\label{eq:deltarhobound}
    \mathrm{Res} = \frac{\sum_n M_n \abs{c_n}^2}{\norm{\rho^{\mathrm{rb}}}^2} \geq \frac{\Gamma \norm{\delta\rho}^2}{\norm{\rho^{\mathrm{rb}}}^2} ,\
\end{align}
implying an upper bound for $\vert \delta\rho \rrangle$. For an observable, $O$, the error in the expectation value resulting from the error in the reduced basis is given by 
\begin{align}
    \delta\expval{O} \equiv \llangle O \vert \rho^{\mathrm{rb}}\rrangle  - \llangle O \vert \rho^{\mathrm{ex}}\rrangle = -(1-c_0) \llangle O \vert \rho^{\mathrm{ex}}\rrangle + \llangle O \vert \delta \rho \rrangle .\
\end{align}
using the triangle and Cauchy-Schwartz inequalities, we can then upper bound this as
\begin{align}\label{eq:obsineq}
    \abs{\delta \expval{O}}  &\leq \abs{\llangle \mathds 1 \vert \delta \rho\rrangle \llangle O \vert \rho^{\mathrm{ex}}\rangle} + \abs{\llangle O \vert \delta \rho \rrangle }\notag\\
    &\leq \norm{O} \left(\abs{\llangle \mathds 1 \vert \delta \rho \rrangle } + \norm{\delta \rho}\right) \notag \\
    & \leq \norm{O}\norm{\mathcal \delta \rho}(\norm{\mathds 1} + 1)\notag\\
    &= \norm{O}\norm{\delta \rho }(\sqrt{d_H} + 1),\
\end{align}
where we also used that $\norm{\rho_{\mathrm{ex}}}^2 = \Tr{\rho_{\mathrm{ex}}^2}\leq 1$ and  $\norm{\mathds 1}^2 = \Tr{\mathds 1 } = d_{\mathcal H}$. Combining inequalities~\eqref{eq:obsineq} and~\eqref{eq:deltarhobound} we obtain the bound error bound on expectation values written in the main text.

\section{Reduced Basis Method for decaying modes.} 

In the following we generalize the iterative reduced basis construction beyond to also include the decaying modes, i.e. the eigenstates of $\mathcal L(\xi)$ with eigenvalues $\lambda\neq 0$. After $N_{i}$ iteration steps, let $\xi_l$ for $l=0,1,..N_i$ be the parameter points where $\mathcal L(\xi)$ has been solved to obtain the $N_s$ states, $\vert \rho_{n}(\xi_l)\rrangle$, with highest real eigenvalues. These are arranged so the real part of the eigenvalues are in descending order, i.e. $0=-\mathrm{Re}{\lambda_0}(\xi_l)\leq -\mathrm{Re}{\lambda_1}(\xi_l)\leq ..\leq-\mathrm{Re}{\lambda_{N_s}}(\xi_l)$. For each point in the full parameter space, $\xi$, we now seek the reduced basis density matrix 
\begin{align}
    \rho^{\mathrm{RB}}_n(\xi)=\sum_{l m} \alpha_{n;ml}(\xi)\rho_{m}(\xi_{l})  ,\
\end{align}
that minimizes the residual 
\begin{align}\label{eq:resn}
    \text{Res}_{n}(\xi)=
    \sum_{\mu\nu}
    \alpha^*_{n;\mu}\big[ &
        (\mathcal{L}^\dag \mathcal{L})_{\mu\nu}(\xi) 
        - \lambda_n^* \mathcal L_{\mu\nu}(\xi)\notag  \\
        &-\lambda_n \mathcal L ^\dag_{\mu\nu}(\xi)  
        + \abs{\lambda_n}^2 S_{\mu\nu} 
    \big]\alpha_{n;\nu}  ,
\end{align}
where $\mu,\nu=(l,n)$ are the indices over both the sample points and eigenvalues, i.e. $\vert \rho_\mu\rrangle = \vert \rho_n(\xi_l)\rrangle$ and $S_{\mu\nu}=\llangle \rho_{\mu}\vert \rho_{\nu}\rrangle$ is the overlap matrix. The above is an expectation value of the positive semi-definite matrix valued function of $\lambda$, $(\mathcal L^\dag(\xi) - \lambda)(\mathcal L(\xi) - \lambda)$, and we therefore have $\mathrm{Res}_n(\xi) \geq 0 $ where this becomes an equality if and only if the exact solution lies completely in the span of the reduced basis. Taking the variation with respect to $\lambda_{n}^*$ and $\alpha_{n;\mu}^*$ we obtain the generalized eigenvalue problem
\begin{align}\label{eq:generalizedeigenvalueproblem}
    \sum_{\nu}\mathcal L_{\mu\nu}(\xi)\alpha_{n;\nu}= \lambda_n \sum_{\nu} S_{\mu\nu}\alpha_{n;\nu} .\
\end{align}
In the case where $S$ is invertible, this turns into a regular eigenvalue problem. However, if many of the reduced basis vectors are almost parallel this can cause the matrix $S$ to be nearly singular (indicated by a high condition number). This is equivalent to the rank of $S$ being smaller than the dimension ($N_s N_i$) reflecting a redundancy in the basis vectors~\cite{Herbst2022Apr}. To circumvent this, we diagonalize $S$ and use this to construct an orthogonal basis, $\vert \tilde{\rho}_\mu\rrangle$, with dimension less than or equal to the dimension of $S$, $d_{\rm rb}\leq N_sN_i$. With this construction, Eq.~\eqref{eq:generalizedeigenvalueproblem} written in terms of the new $(d_{\rm rb}\times d_{\rm rb})$ matrices $\tilde{S}_{\mu\nu}=\llangle \tilde{\rho}_\mu \vert \tilde{\rho}_\nu\rrangle$ and $\tilde{\mathcal L}_{\mu\nu}(\xi)=\llangle \tilde{\rho}_\mu \vert \mathcal L(\xi) \vert \tilde{\rho}_\nu\rrangle$ form a well posed generalized eigenvalue problem with $ d_{\rm rb}$ solutions. For each parameter point $\xi$, we solve Eq.~\eqref{eq:generalizedeigenvalueproblem} and then select the $N_s$ eigenvalues, $\lambda_n$, and eigenvectors, $\alpha_{n;\mu}$, with the smallest values of $\mathrm{Res}_n(\xi)$. Following the approach for steady-states, we find the next sample point by maximizing the combined residual $\mathrm{Res}(\xi)=\sum_n \mathrm{Res}_n(\xi)$ over the parameter space, where the sum is over the $N_s$ smallest residuals for each parameter point $\xi$. We note that while this version of the RBM could also be applied when considering the steady state only, in practice we found that this method introduces larger errors than the one introduced in the main text.  \\
\bibliography{bibliography}